\documentclass[aps,letterpaper,twocolumn]{revtex4}

\usepackage{graphicx}
\usepackage[colorlinks=true, pdfstartview=FitV, linkcolor=blue, citecolor=blue, urlcolor=blue]{hyperref}

\newcommand{\be}{\begin{equation}}
\newcommand{\ee}{\end{equation}}
\newcommand{\bea}{\begin{eqnarray}}
\newcommand{\eea}{\end{eqnarray}}

\begin{document}

\title{
Computational Physics and Reality: Looking for Some Overlap at the Blacksmith Shop
}

\author{Nathan Moore}
\email{nmoore@winona.edu}
\affiliation{
	Department of Physics, 
	Winona State University
	Winona, MN 55987, USA
}

\author{Nicole Schoolmeesters}
\affiliation{
	Winona State University 
	Winona, MN 55987, USA
}

\date{\today}

\begin{abstract} 
The paper describes two general problems encountered in computational assignments at the introductory level.  
First, novice students often treat computer code as almost magic incantations, and like novices in many fields,
have trouble creating new algorithms or procedures to solve novel problems.  
Second, the nature of computational studies often means that the results generated are interpreted via
theoretically devised quantities, which may not meet a student's internal standards for proof when compared to an
experimental measurement. 

The paper then offers a lab/programming assignment, used in a calculus-based physics course, which was devised to address these problems.
In the assignment, students created a computational model of the thermal energy transfer involved in heating an iron rod in a blacksmith's forge.  After creating the simulation, students attended a blacksmithing seminar and had a chance to work with iron and take data on its heating in a coke-fueled forge.  On their return to campus, students revised their computational models in light of their experimental data.
\end{abstract}

\maketitle

\section{introduction}

Computational Physics is quickly becoming a standard component of many departments' degree program and the utility of the field cannot be denied.  It's hard to imagine a modern physics lab without a computer to perform data analysis, and more than just making the task more bearable, most present experimental programs would be impossible without computational automation.\cite{computing_changed_physics_courses,ajp_2008}  
%
 
In addition, the ability to create a working model of a ``real'' system by distilling out the essential physical/mathematical relationships, and then implementing these relationships in a way that captures part of nature is an existential thrill for many students \cite{redish_muppet, chabay_sherwood_computation_in_intro,Gould_atlanta}.  When we allow our students to tackle novel problems and discover real truths on their own, via a simulation they create, we recruit new physics majors \cite{Austin_Peay,Univ_st_thomas}.




\subsection{Novice and Expert Behavior}

When learning how to program, students often seem to grab hold of a certain example program and then treat the syntax therein as a magic incantation.  So long as they don't deviate too far from the example code, their program should compile, run, and produce something close to the desired result.  In so doing, students miss the actual structure of the computer language and the ability to have a computer do ``whatever you want'' rather than ``only the things I have examples for'' \cite{camel_has_two_humps}.  

This is not a trait specific to computational physics, or even computer programming in general.  
Most novices in any field start the learning process by copying the actions of an expert.  In physics, this often occurs in homework problems where students are sometimes asked to  parrot the professor's class examples in different contexts.  
In the same vein, an infant of about six month's age, when seated at the dinner table, will start opening and closing their mouth, just like their parents do when they lift a utensil to their mouth.  

A more sophisticated student will make a plan for how the program should look and function before actually writing the code.  Instead of looking through their library of example programs for one that does something close to the job at hand, an expert will imagine the algorithmic steps necessary to solve the problem, and then figure out what computational machinery (variables, operations, and syntax) best suit the algorithm.  \cite{Phys_educ_1977}

This is not a new division of skill.  
The novice in introductory physics looks at a test problem about the flight of a baseball and tries to figure out what worked class example best matches the problem, perhaps picking the worked trebuchet example from class.  
An expert, working the same problem, will likely exhibit clustering or ``chunking'' of knowledge\cite{chunking}.  Rather than thinking about specific worked solutions, experts tend to think of all the pieces of physics related to the question. 
In this case, the expert might call to mind kinematics, energy conservation, air resistance, and Newton's second law.  
Although the expert's solution might take longer to produce (and contain extraneous details), the fact that the problem is novel is not a insurmountable difficulty for an expert,  \cite{Larkin}.
By contrast, a novice can generally only solve problems that are in their library of solutions.  

The ``best ways'' to move students to the expert state are legion\cite{ajp_resource_letter,UMN_ta_training,Hake}, and there's no utility in repeating the list here.  In the context of programming though, a good start towards creating expert behavior is to require it \cite{incremental_approach,Heller_AJP_1, Heller_AJP_2}.  If we assume that the instruction and materials are clear and sufficient, then the first law of teaching, ``If you want students to do something, grade it!''\cite{pat_heller} would seem to apply.  

\subsection{Misplaced Trust in Technology}
Most calculus textbooks contain a section in the first chapter that might as well be called  ``Lies my calculator told me''\cite{lies}.  As most instructors well know, novice students often trust calculators more than their own heads, and this trust is dangerous, particularly because the calculator is only as reliable as the person who designed and programmed it.  Of course, very few physicists see value in requiring students to be able to compute something like $\log(1.9723)$ by hand, rather, the idea is that students should have the habit of thinking critically about the answer a calculator produces.  


As a more concrete example, consider the student who is using a calculator to evaluate the angle $\theta$ produced by the opposite and hypotenuse sides of a triangle. 
If the hypotenuse is $h=1.0$, and the opposite side is $b=0.2,0.4,0.6,0.8,...$ the student will have little trouble with the $\arcsin$ button on their calculator (using the relation $\theta = \arcsin \frac{b}{h}$).  
If however, the student notices a trend in the value of $\theta$, and out of curiosity plugs in the value $b=1.2$, their calculator will fail them.  Of course, a very good calculator might map $\arcsin$ to the complex plane and produce the answer $\theta=1.57-0.62i$, but most calculators would just produce ``ERROR".  A novice student will generally try the computation again, get the same result, and then proclaim that their calculator is ``broken''.  A more sophisticated student would hopefully think about the geometry implied by $b=1.2$, and perhaps generalize that for values of $b>1.0$ the triangle is no longer right and  will return meaningless results for $\theta$.

Speaking generally, expert students think critically about the answers they generate, either by hand, by calculator, or by computer,  and use unexpected answers either as an indication of error, or as an indicator of some deeper sophistication to the problem \cite{Heller_AJP_1,Heller_AJP_2}.  In computational physics this trait is a complicated one to reinforce, because the nature of the systems studied generally means that the systems are too small, large, hot, cold, or expensive to watch experimentally \cite{CSE_assessment}.   The common approach is to monitor some other quantity, like the system's net momentum or total energy, and check to see if the simulated behavior matches with the theoretically postulated result.  

For example, in an idealized simulation of the moon's orbit around the earth, a student can monitor the gravitational potential energy of the moon and the kinetic energy of the moon, and compare this observation with a theoretical conservation law.  A similar measurement of the net momentum of the moon would reveal that this quantity is not conserved, which an advanced student might link to the external gravitational force exerted on the earth by the moon.

The problem with such measurements is that they're the result of theoretical proofs or conjectures, rather than experimental observations.  In the case of most introductory-level students I have worked with, a plot of kinetic energy is not as persuasive as a computer generated temperature that corresponds to an actual thermometer measurement, taken in person.  
Conservation laws are beautiful, but our beginning students don't believe them.  
Many college freshmen are still in the concrete Piagetian stage \cite{McKinnon_intellectual_development}, and accordingly, ``proof'' to an introductory physics student needs to include something a student can touch.

\section{Goal of the present work}

To summarize, in computational physics an ideal problem will have solution which can be compared to ``real'' data which ideally, is collected by the students.  It seems that agreement between student taken data and a student written simulation is more powerful for learning that agreement with ``facts from a book.''  An ideal problem will also avoid the problem of ``solution by finding a suitable example," described above, by requiring a student to translate physics and mathematics into computer code, which forces a student to think about what the mathematics and numerical approximation actually mean (and thus moves a student along the path toward expert thinking).

The rest of this paper describes a computational exercise which is intended to address the two problems described above.  The problem involves a student-written simulation of heat flow within a bar of iron, which the class then compared to an experiment in the field at a blacksmithing workshop.  After returning from the workshop, students modified their simulations to better account for the experimental data they recorded. 
This project was given in the second semester of a University Physics course at Winona State University.  The class was populated mainly with engineering students who had been exposed to programming via the VPython exercises in the ``Matter and Interactions'' \cite{matter_and_Interactions,nmoore_vpython} introductory physics text.  Rather than simply using a computer to determine a numerical answer, this class had an emphasis on trying to learn about the world through computer simulation.  To make this goal more than just a slogan, computer programming assignments  in this class were evaluated with the rubric in appendix \ref{appx:rubric}.


\section{The Blacksmithing Problem}

The specific problem used in class is described in the following section.  The general idea was for students to get more out of a field experience by taking the time to think about and write a numerical model before going on the trip. After returning, the students had a chance to reflect on the validity of their numerical work by comparing computational predictions to actual experimental measurements.  This approach of $prediction \to data~collection\to reflection$ is not novel, \cite{karplus, Zollman}, but real comparisons of simulation and experimental data seem somewhat rare in introductory computational science.

\subsection{Background and Preparation}

With the thought that other faculty may want to use this project in their own classes, the problem used in class follows.

\textit{
Later this semester, we're planning to go to Dream Acres Farm, near Wykoff, MN, for a seminar on blacksmithing. Although the day will likely be fun, going to school is not all about having fun -- as the two semesters have progressed I hope that you've begun to realize that what we study in class and in lab is closely related to the phenomena we see in our lives every day. While there are many things we can think about in the context of our course material while at the farm, what I'd like for all of you to study is the efficiency of the heating process within the forge. 
}

\textit{
To work and shape iron, the metal needs to be heated to temperatures well above the temperature at which green wood combusts. Iron is ÒtemperedÓ, (based on the annealing temperature the metal can have varying degrees of outer hardness, and inner flexibility) at a temperature range of $100-300^{\circ}C$, and forged (malleable because of the temperature and workable with tools) at temperatures of $700-1400^{\circ}C$. Because green wood starts to smoke at about $150^{\circ}C$, it isn't suitable for heating iron to these high temperatures. In fact, if you have a hot bed of coals and throw on a fresh chunk of wood, you'll actually cool the fire until the wood burns down to charcoal. Instead of fresh wood, we need to use a hotter burning fuel, like charcoal, coal, or coke (which is coal with the tar driven off by heat).  There are a number of semi-scientific works available on blacksmithing, see for example, \cite{edge_of_the_anvil}.
}

\textit{
A piece of iron is heated by simply sticking it in a pile of fuel in the forge, and then forcing air through the fuel 
to increase the rate of fuel combustion. When the iron reaches the desired temperature, you remove it, work it, and then, when too cold to work anymore, reheat the piece in the forge again.
}

\textit{
The initial question is fairly simple. When a piece of metal is put into the fire to be heated, how efficient is the transfer of combustion energy into the metal?
}

\textit{
More specifically, please tackle the three following questions:
\begin{enumerate}
\item 
If you take an automotive leaf spring, dimensions $1\frac{1}{4}$ inches by $\frac{1}{4}$ inch by $12$ inches, initially at $10^{\circ}C$, and heat it in the forge until the bottom $4$ inches glow orange or hotter, about how much energy needs to be added to the iron for this to happen?
\item 
If this heating takes about $45$ seconds, what is the power input (per unit area) from the coals in the forge? What minimum amount of charcoal is burned to accomplish this?
\item 
Use this estimate of energy flow rate in a numerical simulation of energy flow within the bar, and predict a surface temperature distribution for the iron after $15$, $30$, and $45$ seconds of heating.
\end{enumerate}
}

A sample solution to the first two questions is provided in Appendix \ref{appx_estimate}, and the theory underlying a numerical heat transfer scheme, which the students used to build a model, is provided in Appendix \ref{appx_numerics}.  

\subsection{Numerical Preparation}
\textit{
There are probably a number of ways to address this problem, it might be easiest at this point to stick a rod of iron (of known dimension and initial temperature) into the coals, heat the iron, and then pull it out and check the temperature profile with our digital thermometer. We should be able to figure out, based on the temperature gradient, how much energy went into the iron, and further, if we measure the change in mass of coal over the forging period, have a reasonable estimate for how much energy was given up in combustion.
}

\textit{
Before we actually go to the farm, I want you to build a computational model of the rod, heating in a pile of coals. When we actually go to the farm, while one person heats the rod, a second person can measure the time necessary for the metal to heat up to forging temperature. A good computational simulation of the process should be able to duplicate these results with reasonable accuracy.
}

\textit{
When we get back from the farm, I'd like you to refine your computational model to account for any discrepancies between the data you took, and the temperature profile you measured experimentally. 
}

\subsection{While at the Farm}
\textit{
While in the forge area, and after you've been instructed about safety and technique, I want you to take a piece of metal stock and score regular marks on it with a file. Then heat up the piece of iron until its quite hot (red-orange is sufficient) noting the clock-time necessary for the piece to heat to this color.
}

\textit{
When heated, remove the iron from the coals and take temperature measurements with the IR thermometer at the marked locations, at $30$ or $60$ second intervals (whatever seems feasible), for at least $10$ minutes.
}

\textit{
Your measurements right after the iron is removed from the fire should correspond closely to the numerical heat transfer simulation you wrote before going on the lab. This experimental data set will allow you to calculate the rate at which heat flows from the coals into the fire (something we had to guess at in the simulation).
}

\subsection{Reflection and Analysis after we return}
\textit{
Given the data you took for the heating and cooling of the iron in the forge, refine the computational model for heat transfer you wrote before the trip. As a write up for the programming project please answer the following questions (including figures, equations, and diagrams as appropriate):
\begin{enumerate}
\item
What average power density does a stoked fire provide to a piece of iron stock? How did you figure this out? How accurate do you think this estimate is? (Include a well-reasoned estimate of the uncertainty in your answer.)
\item 
Did you include radiative losses from the un-heated end of the iron in your model of heat transfer? Do you think they're necessary? Do you think the physics included in the computational model is complete? As always, justify any statements you make with respectable scientific arguments.
\end{enumerate}
}

\section{Student Solution}
\label{solution}

\begin{figure}
\includegraphics[width=\columnwidth]{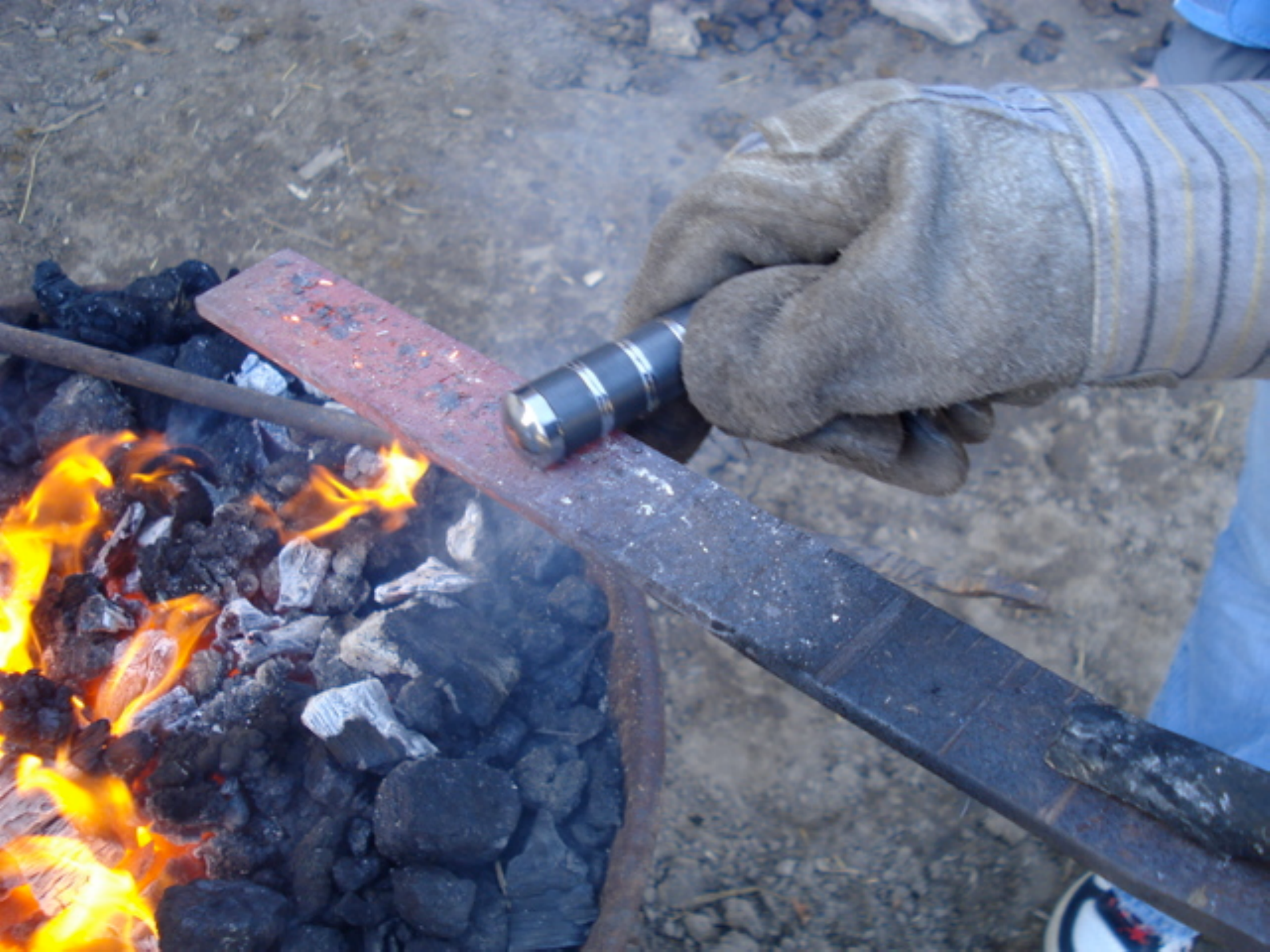}
\caption{
(Color online)
This image shows the basic set up of the lab.  One benefit of the high temperatures involved is that a magnet can be used as a primitive thermometer.  Iron undergoes a ferromagnetic/paramagnetic transition at about $770^{\circ}C$, so by running a magnet down the length of the stock and finding this transition point, students can estimate how much heat has been transferred into the bar.  
For departments without an infrared thermometer, finding this magnetic transition point is a straightforward (and cheap) way to obtain quantitative data to check a simulation.
}
\label{fig:curie_temp}
\end{figure}

The field trip ran roughly as described, and in addition to the temperature measurements described, students were able to see the ferromagnetic/paramagnetic phase transition of iron by heating the bar and then bringing a large ``cow'' magnet near the sample, see figure \ref{fig:curie_temp}.   We also brought a small spectrometer along to the shop and above the blackbody background, we saw clear sodium spectral emission lines at about $580nm$.

The remainder of this section (\ref{solution}) is the solution produced by one student, Nicole Schoolmeesters, who went on the trip.  The code Nicole wrote for the project is available online \cite{nicole_code}.  She writes:

\begin{figure}
\includegraphics[width=\columnwidth]{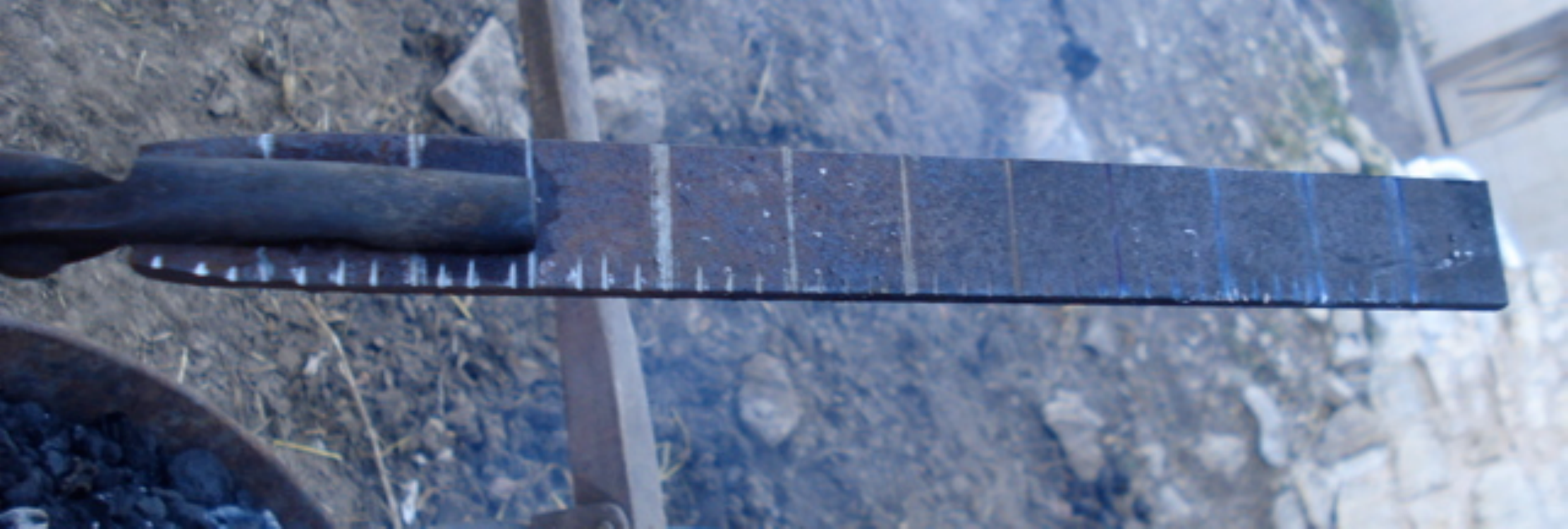}
\caption{
(Color online)
Students created a distance scale on the iron rods by cutting notches into bar stock with a file.  There was a metal tape-measure handy in the blacksmith shop and students felt that inches were a natural distance unit to use in the lab.  The simulations used metric units exclusively, and simulation output was converted into inches to make a comparison with the measured temperature data.
}
\label{fig:distance_markings}
\end{figure}

At a blacksmithing workshop at Dream Acres, we investigated heating iron rods to see first hand the different properties of iron.  For this lab, we worked with an iron rod with the dimensions of $12$ inches by $1 \frac{1}{4}$ inch by $\frac{1}{4}$ inch that is initially at $10^{\circ}C$.  
A narrow notch was scratched across the width of the rod at one inch intervals to reference surface temperature measurements, see figure \ref{fig:distance_markings}.  
For each trial, approximately $4$ inches of the rod was stuck into the hottest part of the coals for certain length of time.  Then we quickly measured the temperature with an IR thermometer at every inch notch mark along the length of the rod.  Between each trial, the rod was quenched so that the initial temperature of it would return to approximately $10^{\circ}C$.  The data recorded at the forge was then  compared to the output data from the model rod in the simulation written before the field trip. 
	
The goal of this lab was to see if we could make our VPython program output temperature values along the surface of the iron rod that duplicates the real life data.  
The computer code simulated $30$ seconds of heating, and the code's output could then be compared to the $15$ second and a $30$ second temperature measurements which we took at the blacksmithing forge.  
My program simulates the heat flow within a rod with the heat equation, which is derived from the ideas of calorimetry and heat conduction, and is contained in Appendix \ref{appx_numerics}.  
When the rod is stuck into the inside of the forge, energy flows into the rod and is seen as an increase of temperature.  
I described this influx of heat as Ò$P_{coals}$Ó, which is the the rate of energy flow per unit area on the outside surface of the rod.  
For the portion not stuck into the coals, I also included in my simulation the non-contact heat flows from the rod to the rod's surroundings via the Stefan-Boltzmann radiation law (for a blackbody, emissivity $1$).


Although the simulation code is written with metric units, we used english units throughout the paper because there was a carpenter's tape measure in the blacksmith shop and inches was a natural unit for the problem.
In the simulation, the rod is subdivided into small chunks to more accurately model how energy flows into, through, and out of the rod.  There are $48$ chunks along the $12$ inch axis, $16$ chunks along the $\frac{1}{4}$ inch axis, and $20$ chunks along the $1\frac{1}{4}$ inch axis.  This means that each ``finite element'' in our model has dimensions of $\frac{1}{4}$ inch by $\frac{1}{32}$ inch by $\frac{1}{16}$ inch.  

Although the material properties of iron change with temperature, in this model we held the following properties constant:  
specific heat, $c_p= 449 \frac{J}{kg~K}$; 
density, $\rho = 7870 \frac{kg}{m^3}$;  
and thermal conductivity, $k= 80.2 \frac{W}{m~K}$.   
The entire system's temperature was initialized to $10^{\circ}$.

The timestep for the program is determined by the numerical stability of the algorithm described in Appendix \ref{appx_numerics}, and was determined largely by trial and error.  
The largest timestep that we used is $dt=0.002$ seconds.
On a $2.16GHz$ MacBook the program then takes about $3$ hours to simulate $30$ seconds of forge heating, \cite{openmp}.

The bulk of the code is derived in Appendix \ref{appx_numerics}, and an implementation in VPython is available as a supplementary file on arxiv.org, \cite{nicole_code}.

\begin{figure}
\includegraphics[width=\columnwidth]{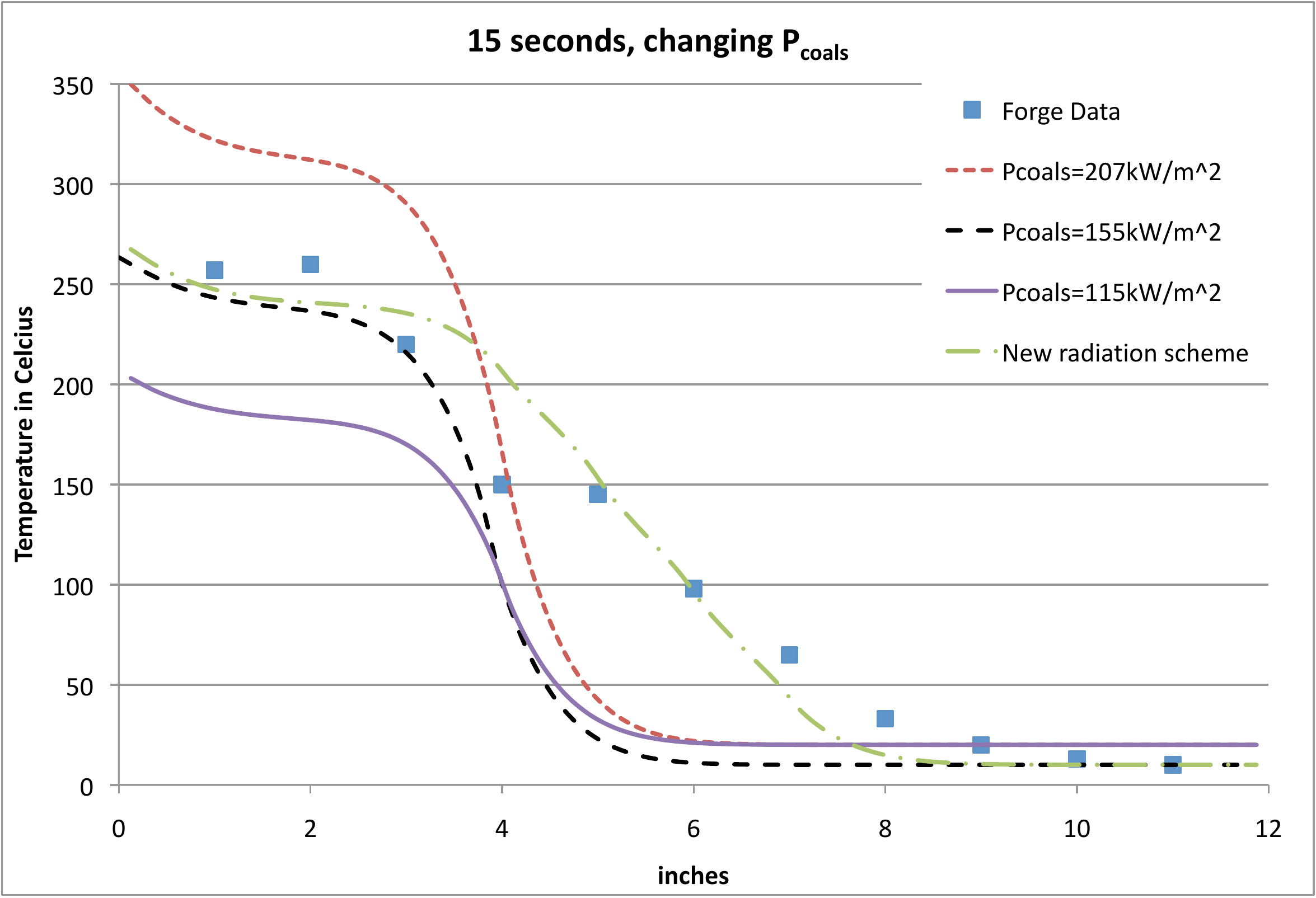}
\caption{
(Color online) Simulation and experimental surface temperature data for the steel (iron) rod shown in earlier figures.  The data was taken after the rod has been heated in the forge for $15$ seconds.
The purple, black, and red lines correspond to the simulation where heat is transferred into the rod from direct contact with the coals.  The green line is from a modified simulation which adds an approximate radiation transfer scheme.  This revised version of the code fits the experimental data (blue boxes) fairly well.  
Surface temperature measurements on the rod were taken with a Fluke 68 IR Thermometer, which the manufacturer clams has an accuracy of $\pm1^{\circ}C$.
}
\label{fig:15_sec}
\end{figure}

\begin{figure}
\includegraphics[width=\columnwidth]{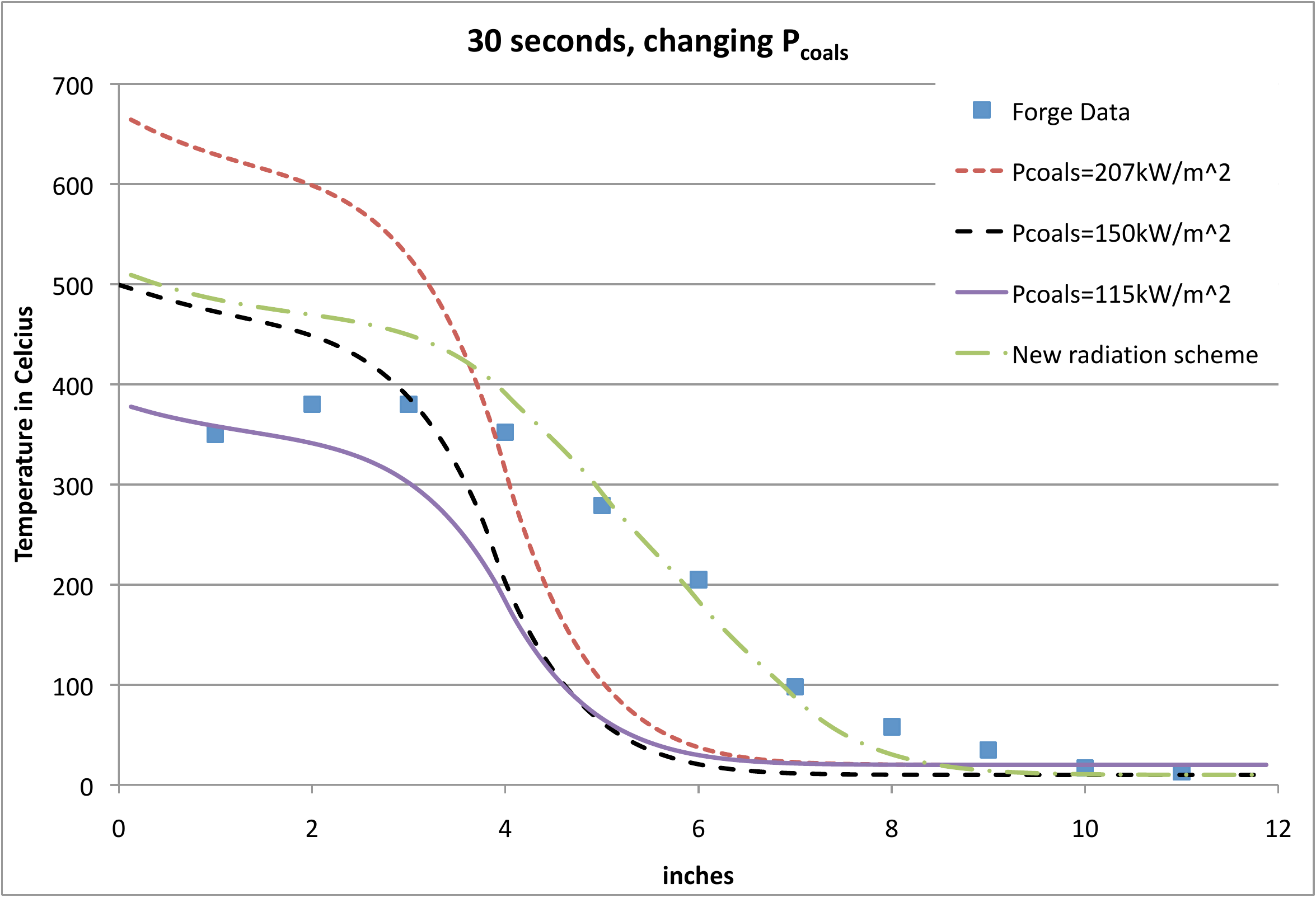}
\caption{
(Color online)
This figure contains the same plotted variables as  figure \ref{fig:15_sec}, but the time corresponds to $30$ seconds of heating in the blacksmith forge.  Note again the poor agreement of simulation to data when the radiation transfer scheme is not included.  Figures 3-6 use a horizontal length scale, along the metal bar, of inches, where $0$ corresponds to the end of the rod in the coals.  The simulation uses metric units, and output was converted to the english length in order to correspond to the measurements taken at the blacksmithing workshop.  
}
\label{fig:30_sec}
\end{figure}

A comparison of the forge temperature data and the prediction values from the simulation written before the trip is shown in figures \ref{fig:15_sec} and \ref{fig:30_sec}.
The calculated value for $P_{coals}$ in the first trial was $\approx 207 kW/m^2$.  As the figures show, the simulation output and the actual forge data have a similar shape for the four inches of the rod stuck into the coals, but the simulation consistently produces higher temperature values for both the $15$ second intervals and the $30$ second intervals.   

To understand why the simulation produced higher temperatures, I ran the simulation again and decreased the value of $P_{coals}$.  The $15$ second forge data suggested $P_{coals}\approx  155kW/m^2$.  
The subsequent run has $15$ second output which is better correlated to the $15$ second forge data, but the simulation output was still too high for the $30$ second data (again, see figures Figures \ref{fig:15_sec} and \ref{fig:30_sec}).  
Accordingly, I decreased the heat input again, only this time using the $30$ second forge data to determine a value of $P_{coals} \approx 115 kW/m^2$.  
This time, the temperature output from VPython was too low for the $15$ second line in that simulation, but the $30$ second line shares a similar shape and is within the expected temperature range when compared to the forge data.  From this set of varied inputs it seems that the rate of energy flow into the rod must decrease as the temperature increases within the rod. 

\begin{figure}
\includegraphics[width=\columnwidth]{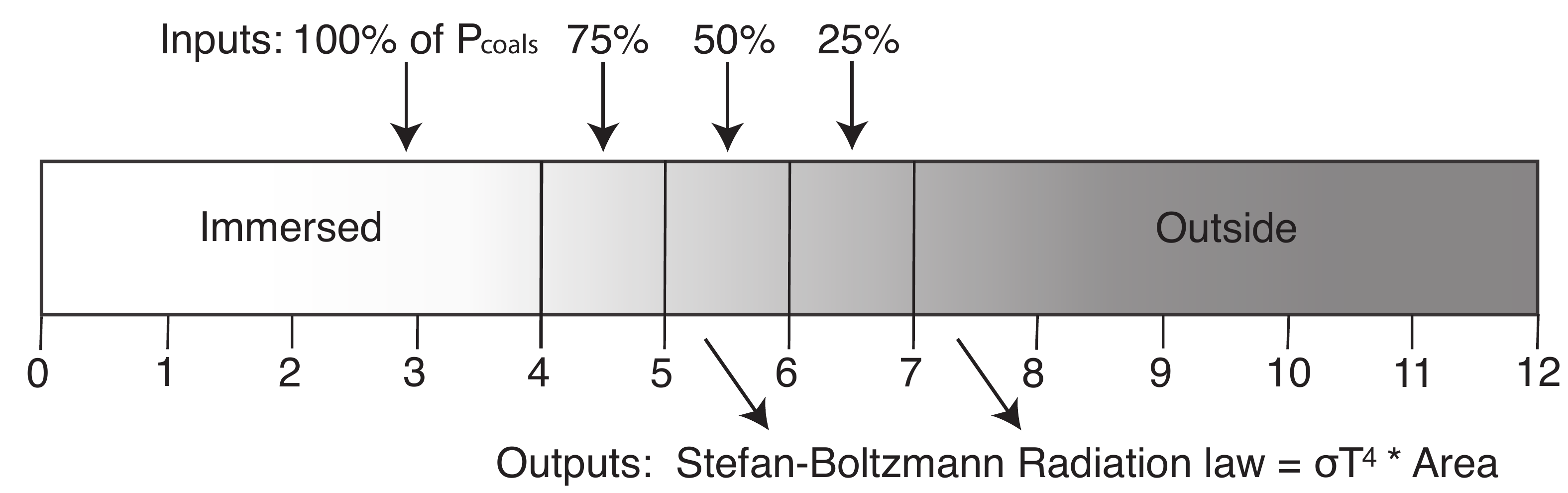}
\caption{
The original model of thermal energy flow into an iron rod placed approximately $4$ inches into a blacksmithing forge has a constant incoming power density, $P_{coals}$.  The modified model decreases $P_{coals}$ by $25\%$ for every inch within the middle portion of the rod, as seen by the gradation of grey between the immersed 4 inches and the cold, outer portion of the rod.  In both cases, the portion of the rod not inserted into the coals radiates heat to the colder surroundings via the Stefan-Boltzman Radiation law. 
}
\label{fig:transfer_model}
\end{figure}

Although the $15$ second forge data and the VPython output where $P_{coals}\approx  155kW/m^2$ match well, another refinement to the VPython program could be made for the middle portion of the rod.  When plotting the temperature data measured at the blacksmithing forge, there is a gradual decrease in temperature between the four inches of the rod stuck into the hottest part of the forge and the four inches at the opposite end of the rod that is unaffected by the heat from the forge.  However, the VPython code was written so that there is an abrupt change to no additional heat inflow past the four inch mark where the rod is immersed into the forge.  The middle portion of the rod, where it is still surrounded by hot coals but not physically touching the coals should have a decreasing amount of energy flow into it, proportional to the distance a segment of the rod is from the surface of the bed of coals.   This observation leads to the heat transfer scheme shown in figure \ref{fig:transfer_model}.

As implemented, this modification decreases $P_{coals}$ by $25\%$ for every inch within this middle area.  For example, between inch 4 and inch 5, the power density is $75\%$ of the original value.  Between inch 5 and inch 6, the power density is half that of the original, and between inch 6 and inch 7 it is $25\%$ that of the original.  The remaining part of the rod beyond inch 7 no longer has any heat flow into it from the coals.  After including this  modification, using $P_{coals}=155kW/m^2$, as figure \ref{fig:15_sec} shows, I found the the simulation output fit the forge data really well for the $15$ second temperature output compared to the previous trials.  As for the $30$ second data, shown in figure \ref{fig:30_sec}, the temperature output from the simulation was high for the first four inches, but fits the rest of the forge data fairly well. 

\begin{figure}
\includegraphics[width=\columnwidth]{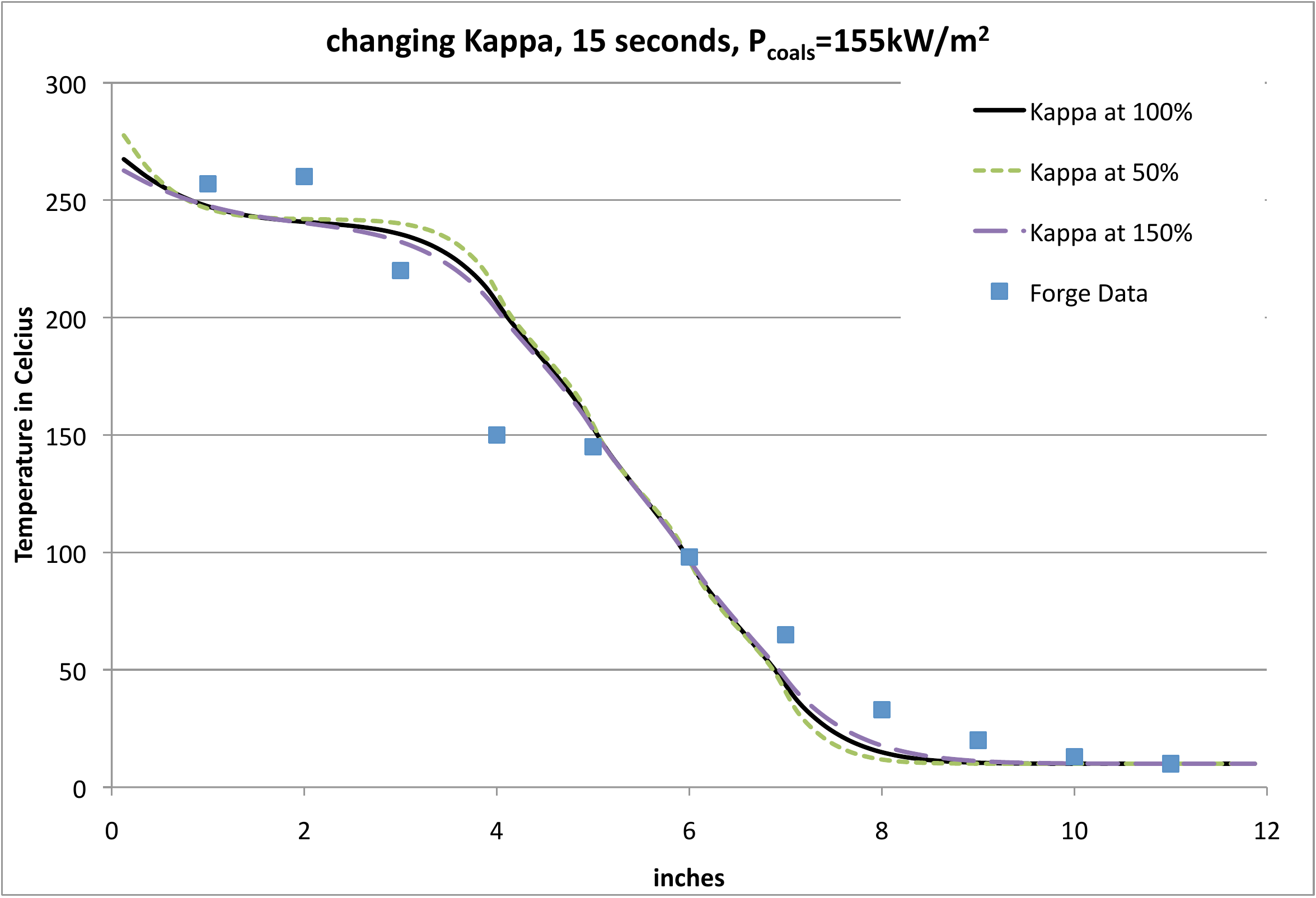}
\caption{
To see if variations in specific heat capacity, density, or thermal conductivity might account for the discrepancy between the original simulation and the experimental data, we re-ran the simulation (with the radiation scheme included) with wildly divergent $\kappa$ values of the original $\kappa_0\approx2.27\times10^{-5}\frac{m^2}{s}$, $\kappa=1.5\kappa_0$, and $\kappa=0.5\kappa_0$.  As the output indicates, variations in $\kappa$ will not significantly affect the simulation output.}
\label{fig:vary_kappa}
\end{figure}

The thermal conductivity($k$), specific heat ($c_p$), and density ($\rho$) of iron change with temperature.  Additionally, as a piece of iron is worked, the surface composition can change subtlety.  Given this, I wondered if a different value of $\kappa=\frac{k}{c_p \rho}$, the coupling constant in the heat transfer equation, might also account for the discrepancy seen in the comparison of experimental data and the original simulation.  (Basically, I wanted to see if the radiation transfer scheme described above was really necessary, or if variations in $\kappa$ might lead to the same surface temperature distribution.)  To test this idea, I ran the simulation again with a drastically increased and decreased value for $\kappa$ to see if there was a significant change to the shape of the graphs.  As seen in figure \ref{fig:vary_kappa}, subtle variations in $\kappa$ cannot account for the differences seen between the temperatures measured at the forge and the temperatures produced by the simulation.

\section{Alternative implementations}

While few campuses have blacksmith shops in their science complex \cite{blacksmithing_exceptions}, the direct comparison of simulation with experimental data is a powerful and necessary thing for an undergraduate in science to see and experience.  The data taken in this project is used to check the accuracy of the simulation and hopefully suggest other physics.  Many other variations of this lab, not requiring a blacksmith's forge, are feasible and we now discuss a few options.

The simulation described in Appendix  \ref{appx_numerics} accounts for heat transfer within a box-shaped object with known power flux through the surface of the object.  To simulate the forge, an instructor could use somthing a simple as an electric toaster oven or a ``Weber'' style charcoal grill, both of which would have power densities sufficient to heat a piece of metal stock.  

Although not discussed, the simulation should also model quenching of a hot metal object fairly well (although we have not taken data to verify this).  Students could pre-heat a piece of iron in a water bath and then quench the stock in ice water or liquid nitrogen.  In both cases, the fundamental physics should be similar, although a temperature dependent heat flux may be necessary if the object is quenched for a long time.  
   
As discussed in the caption to figure \ref{fig:curie_temp}, iron undergoes a paramagnetic/ferromagnetic phase transition at about $770^{\circ}C$, so in lieu of an IR thermometer, a simple bar magnet can be used as a cheap temperature probe by locating the point at which magnetism ``goes away'' and then tracking the position of this transition point as the bar is heated for different time intervals.   Alternatively, students can use digital cameras to record the color distribution across a heated bar (as illustrated in figure \ref{fig:curie_temp}), and then compare these images with the forging color scales availabe in blacksmithing books \cite{swedish_blacksmithing}.  Of course, these color profiles are directly related to the blackbody distibution, and the opportunity for another learning cycle is again available.

Finally, the magnetic phase transition described above makes a compelling follow-up project for interested students.  Students can study the boltzmann factor (see \cite{matter_and_Interactions}, chapter 11 for a excellent intro-level derivation), and then create an Ising/Monte-Carlo style model of the iron stock which would give a qualitative picture of the phase transition described above.

\section{Conclusions}


Computational physics is a natural part of a learning cycle in physics, not only because it illustrates concepts which are hard to imagine, but also because comparisons of simulation to data can suggest (or demand) new phenomena.  This paper describes such a problem, in which radiative transfer, initially neglected, needed to be included for the simulation to better model reality. The idea that data demands a better description of nature is the essence both of physics and inquiry.

This problem could be adapted to other campuses by using a different heat source or sink in place of the blacksmithing forge.  For example, the power from the forge could be simulated by placing metal stock in a toaster oven.  Similarly, the quenching process could be simulated by dipping stock, initially at room temperature, in a bath of ice-water or liquid nitrogen.  

In our view, the implementation details are less important than the feedback loop described explicitly in this work.  The greatest beauty of physics is that it is fundamentally a \text{real} description of the world.  When creating computational models, students need to be reminded and persuaded of this connection to reality by evaluating their solutions with data sets which are as tangible as possible.

\begin{acknowledgments} 

Nathan Moore taught the section of Physics 222 at Winona State University in Fall 2008 in which this problem was implemented.  He came up with the problem, wrote the introduction to the paper, and contributed Appendix \ref{appx_numerics}.

Nicole Schoolmeesters was a student in Moore's University Physics course. Her VPython program was used to generate the figures shown in the paper, and she took the experimental data plotted in figures \ref{fig:15_sec},  \ref{fig:30_sec}, and  \ref{fig:vary_kappa}.  She came up with and implemented the additional radiation heat transfer mechanism described in section \ref{solution}.

This work was supported in part by a Minnesota State Colleges and Universities ``Learning Games and Simulations'' grant.  The authors also wish to thank Winona State University for their support of such an unconventional physics field trip.  Additional thanks is due to Jeff Anderson of Winona State's Math department, who clarified a few issues related to convergence criteria for numerical solutions to the  3-d Laplace equation.

The grading rubric provided in appendix \ref{appx:rubric} was developed by Andrew Ferstl and Nathan Moore, both of Winona State University.

Finally, the authors thank Todd Juzwiak of DreamAcres Farm, Wykoff, MN, who was kind enough to open up his farm and blacksmithing shop to the class. 
\end{acknowledgments}


\appendix

\section{Grading Rubric for Programming Problems}
\label{appx:rubric}

\textbf{Rubric for Assessing student understanding of basic physics (Using computer modeling)}

This is the rubric that we will use to evaluate your computer programs. Of course, at a
minimum your program must work, but to get a good grade for the assignment, 
you must adequately
document your program with comments that explain what each part of the program is
doing, explain why you are solving the problem in the way you chose, and explain the relevant 
physics . 
Some of these comments should be in the body of the code, and some of the
documentation (like plots of the results, derivations, etc) should be in an attached report.
As reflected in this rubric, you will be graded for not only how well your program works but
how well you have articulated your solution/program/physical-model to the grader.

\begin{figure}
\includegraphics[width=\columnwidth]{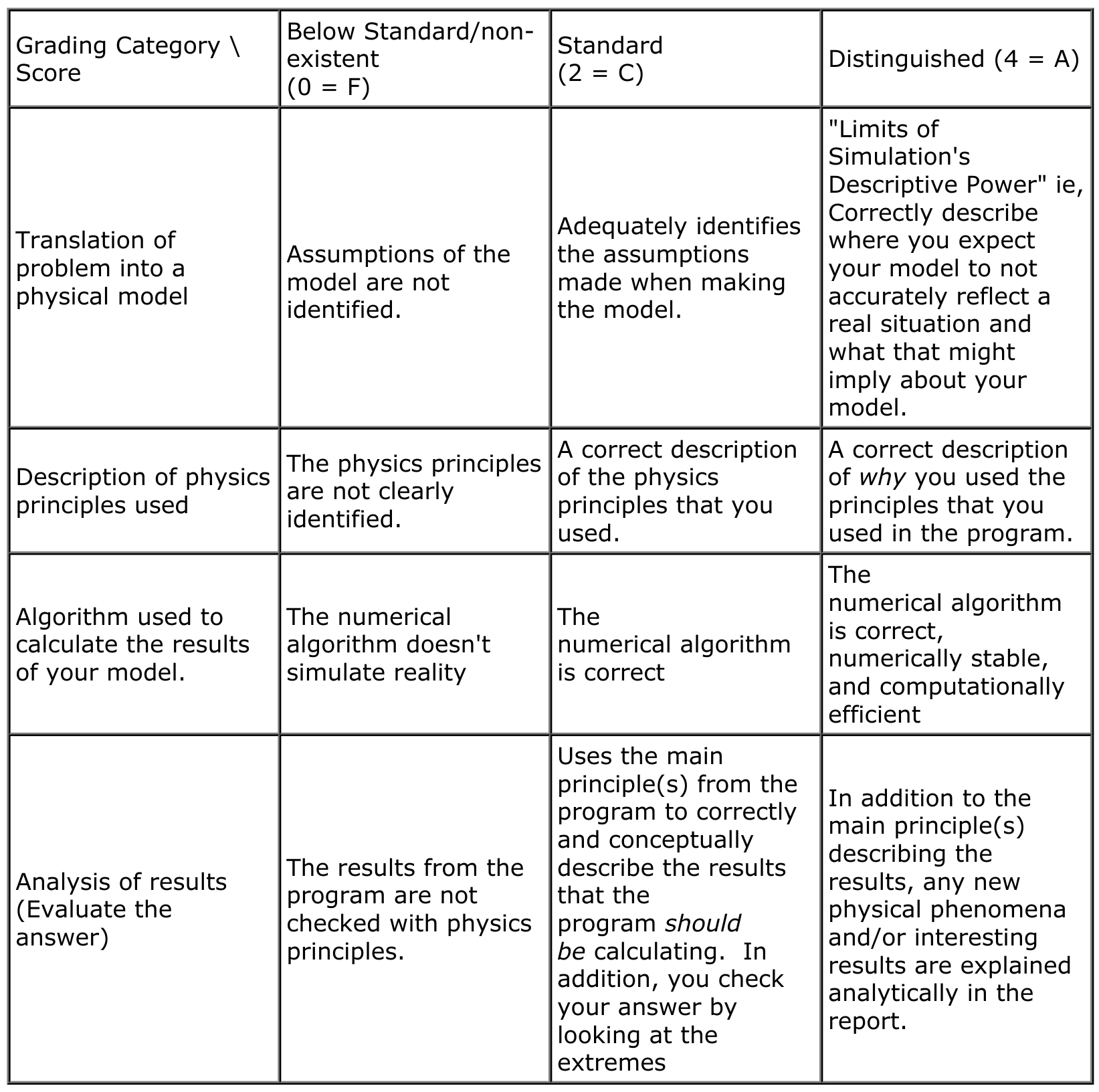}
\caption{This rubric was developed by Andrew Ferstl and Nathan Moore when they started including VPython programming problems in their introductory physics classes.  
Rather than simply using a computer to generate a number, we emphasized in class the importance of trying to recognize and understand the physics that a computer model illustrates.  In order to change the way that our students think about programming assignments, we gave them the following categories to guide their perspective, code, and solutions.
}
\label{fig:rubric}
\end{figure}

\section{Heat flow estimates}
\label{appx_estimate}

\textbf{Question:}
If you take a leaf spring, dimensions $1\frac{1}{4}$ inches by $\frac{1}{4}$ inch by $12$ inches, initially at $10^{\circ}C$, and heat it in the forge until the bottom $4$ inches glow orange or hotter, about how much energy needs to be added to the iron for this to happen?

\textbf{Solution:}
The calorimetry equation, $dU = m c dT$,  can be used to determine the amount of energy needed to heat the first four inches of the rod.  The mass is found by multiplying the volume of the rod that was heated ($4 \times 1\frac{1}{4} \times\frac{1}{4}$) by the density of iron.  The approximate temperature of an iron rod that glows orange is $1000^{\circ}C$, so the change in temperature that the rod undergoes is about $1000^{\circ}C$.  The amount of energy that is then needed to heat up the rod is the product of mass, specific heat, and the temperature change, about $74kJ$.  

\textbf{Question:}
If this heating takes about $45$ seconds, what is the power input (per unit area) from the coals in the forge? What minimum amount of charcoal is burned to accomplish this?

\textbf{Solution:}
If heated for $45$ seconds, the power input from the coals can be calculated with the relation,  $\frac{dU}{dt} = P_{coals} \times Surface~Area$.   This works out to be $P_{coals}\approx 207 \frac{kW}{m^2}$. 

The energy density of coke is about $29.6 MJ/kg$ \cite{efunda_iron} and the energy needed to heat up the rod is $74kJ$.  Therefore, if \textbf{all} of the energy from the coke goes into the iron you'd need about $~0.3kg$ of coke.  Of course, this is a dramatic underestimate.

\section{Creating a Numerical Model}
\label{appx_numerics}

This appendix contains a derivation of the numerical method that students used to create a computational model for heat flow within a cubical object.  

To build a useful model, we need to understand the system fairly well.  Lets assume that we're trying to build a wood chisel out of a leaf spring from a 1970's era full-size pickup truck.  This iron, (technically, steel), is high carbon, perhaps as high as $1.5\%$, and great for making tools.  The piece of metal is roughly $2$ inches wide, $16$ inches long, and about $1/4$ inch thick.  To make the stock malleable, we need to heat  the bottom $4$ inches of the spring by burying it in the pile of coke (or charcoal) in the forge.  The question we want to answer is ``If one end of the bar is stuck into the coals, after a $15$ or $20$ second heating period, what temperature distribution should we see along the length of the bar?"

\subsection{Necessary Basic Ideas}
We first need to remember a few basic relationships from thermodynamics  to be able to explain the mechanism for heat transfer within a material.  The rate of heat flow, $Q$, through a barrier like the  exterior wall of your house, is related to the temperature gradient between the inside and outside of the house wall, $\Delta T$, the area of the wall, $A$, and the thickness of the wall, $\Delta x$, by the equation, 
\be
\frac{\Delta Q}{\Delta t}=-\frac{kA\Delta T}{\Delta x}.
\ee

In this equation, $k$ is the ``thermal conductivity'' of the barrier.  For air, $k=0.025\frac{W}{m ~K}$, which is quite low, and should make sense, as still air does not conduct heat well.  Iron has a conductivity of $k\approx50-80\frac{W}{m ~K}$\cite{efunda_iron}, depending on temperature and composition.  By contrast, copper has a $k$ value of nearly $400~\frac{W}{m ~K}$, which can also be explained in terms of common experience.  

Metal objects often feel cold because the high thermal conductivity of the material allows lots of heat to conduct away from your skin, which creates the sensation of ``cold".  If you touch comparable sized pieces of iron and copper, it is likely that the copper will feel colder because if its greater thermal conductivity.  On a hand-waving level, copper's greater conductivity occurs  because there are more ways for energy to move within a metal, specifically electrical and acoustical conduction.  
Naturally, there is also a differential form of the heat conduction law which is more suited to our purposes in this project,
\be
\frac{dQ}{dt}=-k A \frac{dT}{dx}.
\label{eq:heat_cond}
\ee

The relationship between the increase in internal energy of a material and the observed temperature change of the same material is called calorimetry.  Formally, the definition relies on ``heat capacity'', $C$, which is the rate of change of internal energy, $U$, with respect to temperature $T$.  If we talk about the change in temperature and internal energy of a specific mass $m$ of material, subject to a constant (atmospheric) pressure, the rate of change is called ``specific heat'', $c_p$.  The useful relation is then, 
\be
\Delta U = m c_p \Delta T,
\label{eq:calorimetry}
\ee
where the change in internal energy $\Delta U$ causes a change in temperature $\Delta T$.  Most metals have a specific heat of approximately $3R$ per mole, a beautiful result from the Einstein Solid model. For iron, the specific value is $c_p=449\frac{J}{kg~K}$ at room temperature, but there are  significant changes in this value as the  the iron's temperature changes.

\subsection{How does the heat actually move in the solid?}

To model the heat flow, I suggest you use the classical heat equation, which can be derived by combining the previous results for calorimetry and conductivity.   The derivation of this equation comes by using calculus to combine the formulas for energy storage (Calorimetry) and energy flow (Thermal Conduction).

First, think about a small cubical chunk of material.  The cube has density $\rho$, volume, 
$V = \Delta x \Delta y \Delta z$, and mass, $m=\rho \Delta x \Delta y \Delta z$.  At a certain point in time (assuming constant heat capacity), the change in internal energy of the box can be related to the change in the temperature of the box by the differential form of equation \ref{eq:calorimetry} above,

\be
dU = c_p \rho V dT 
\ee  

In general, if the heat flow into or out of the box is described as Q, we can go a step further and say,

\be
\frac{d Q}{dt} = \frac{dU}{dt} = c \rho V \frac{dT}{dt}
\label{eq:diff_cal}
\ee

Now, we can specify the heat flow into the box by thinking about the heat conduction equation described above.  Specifically, if we think about ONLY the x direction, and say that the box is centered on x, then there will be two relevant heat flows, one from the $x-\frac{\Delta x}{2}$ side, and another from the $x+\frac{\Delta x}{2}$ side.  Mathematically then, the net heat flow is the difference between what flows in and what flows out:

\be
 \frac{dU}{dt}=Q|_{x-\frac{\Delta x}{2}} - Q|_{x+\frac{\Delta x}{2}}
 \ee

We know the formula for heat conduction, equation \ref{eq:heat_cond}, so the task now is plug in that relation and see the math that falls out.  For the geometry we're using, $A = \Delta y \Delta z$

\bea
\frac{dU}{dt} &=& k A \left(\frac{dT}{dx}|_{x+\frac{\Delta x}{2}} - \frac{dT}{dx}|_{x-\frac{\Delta x}{2}}\right)
 \label{eq:cond_2}
 \eea

Then, if we combine the two equations for $\frac{dU}{dt}$, \ref{eq:heat_cond} and \ref{eq:cond_2}, there's a beautiful result:

\bea 
\frac{dT}{dt} &=& \frac{k A}{c_p \rho V}\left(\frac{dT}{dx}|_{x+\frac{\Delta x}{2}} - \frac{dT}{dx}|_{x-\frac{\Delta x}{2}}\right).
\eea

If you think about the definitions, $\frac{A}{V}=\frac{\Delta y \Delta z}{\Delta x \Delta y \Delta z}=\frac{1}{\Delta x}$, and then, using the fundamental definition of a derivative, 

\be
\frac{dT}{dt} = \frac{k}{c_p \rho}\frac{d^2 T}{dx^2}.
\ee

In three dimensions the arguments are the same (replace x with y, etc), and the constant $\frac{k}{c_p \rho}=\kappa$, and the result becomes:

\be
\frac{\partial T}{\partial t} = \kappa \left(\frac{\partial^2 T}{\partial x^2}+\frac{\partial^2 T}{\partial y^2}+\frac{\partial^2 T}{\partial z^2}\right).
\label{eq:heatflow}
\ee

This final result is a ``partial differential equation", which can be solved analytically (ie using math to figure out a closed-form solution) for very simple geometries, and numerically for more complex (real) geometries and heat flows.  It is a virtual certainty that many (or nearly all) of the appliances you plug into the wall are thermally simulated before being actually built.  Obvious examples are CPU heat sinks, the cooling coils in refrigerators, the heat exchanger in a furnace, the engine bay of your car, etc.

\subsection{A Numerical Model of the Bar }

How do we use a computer to solve this problem?  First, we'll have to think about some sort of array-type representation of the iron bar.  One of the common approaches is to think about the material as consisting of little pieces, each with an individual temperature, conductivity, mass, etc.  Given the size of Avagadro's number, we can't make the chunks of iron the size of an atom, or even a few atoms, but if the chunks are much smaller than the object itself, the results can be quite accurate.  More specifically, we want the chunks of iron to be smaller than the spatial scale of the phenomena we're interested in studying.  In this case, since the part of the bar plunged into the coals is about 4 inches long, we should use a chunk size that's much smaller than 4 inches.  The obvious limit to this reasoning is that as the chunk size decreases, the computation time required for a simulation increases in proportion to the number of chunks.  Lets assume that the bar has dimension $L_x \times L_y \times L_z$, where $L_x=16 in$, $L_y=1/4 in$, and $L_z=2 in$.  Further, lets say that all chunks have dimension $\Delta x \times \Delta y \times \Delta z$.  You can imagine then that the number of chunks in the x direction is $N_x=L_x/\Delta x$ and so on.

With this discretization then, we can talk about a representation in computer code.  Since the differential equation for heat flow is written in terms of the temperature, T, we can represent the continuous distribution of temperature as a 3-d array, \texttt{temp[i][j][k]}, where the indices run from $i=0,N_x-1$, $j=0,N_y-1$, and $k=0,Nz-1$.   


Once the array of temperatures is defined, we can work with an individual array element by referencing it by its 3-d location in the lattice of mass chunks.  If we define the $i=0$ chunk to span from $x=0$ to $x=dx$, and the $i=N_x-1$ chunk to span from $x = L_x-dx$ to $x=L_x$, then we can initialize the bar to have an initial temperature with an iteration over the whole array.  The mechanics of this are specific to the computer language used, see the associated example program for an implementation in Visual Python.




Remember, to use a computer to model the heat flow, we need to assume that the bar of iron is made up of little boxes, each with a temperature of  \texttt{temp[i][j][k]}.  The trick now is to re-write the equation for heat flow, equation \ref{eq:heatflow}, to accommodate this numerical approximation of the iron, rather than the continuous distribution that the equation above describes.



\subsection{Finite Differences}

This approach is inspired by a discussion of numerical solutions to electrostatics in Chapter 5 of ``Computational Physics'' by Giordano and Nakanishi\cite{Computational_Physics}.
Although the physical context is quite different, similar mathematical techniques can be used for both problems. 

To implement a computational solution, we need to figure out how to represent $\frac{\partial^2 T}{\partial x^2}$ in terms of \texttt{temp[i][j][k]} in the x, y, and z directions.  To do this, you need to think about approximating this second derivative as a small box to box change, or ``finite difference''.  Specifically, if we want to know the spatial derivative of temperature, $\frac{dT}{dx}$, we can say that,

\be
\frac{dT}{dx} \approx \frac{\Delta T}{\Delta x} = \frac{T(x+\Delta x)-T(x)}{\Delta x}
\ee

This approximation 
increases in accuracy as $\Delta x$ gets smaller.  With a little creative re-ordering of the x-axis, we can say that the derivative at $x=x_0$ is,

\be
\frac{dT}{dx}\Big |_{x=x_0} \approx \frac{T(x_0+\frac{1}{2}\Delta x)-T(x_0-\frac{1}{2}\Delta x)}{\Delta x}.
\ee

Now, if we want to describe the second derivative at $x=x_0$, we just have to (creatively) say that the second derivative is the difference between two first derivatives, and pick the first derivatives to occur at $x=x_0\pm\frac{1}{2}\Delta x$.  Mathematics should make this more clear:

First, the second derivative, expressed in terms of first derivatives:

\be
\frac{d^2 T}{dx^2}\Big |_{x=x_0} \approx  \frac{1}{\Delta x}\left( 
  \frac{dT}{dx} \Big |_{x=x_0+\frac{1}{2}\Delta x} 
  - \frac{dT}{dx} \Big |_{x=x_0-\frac{1}{2}\Delta x}
\right).
\ee

Next, the two first derivatives, evaluated at $x=x_0\pm\frac{1}{2}\Delta x$

\bea
\frac{dT}{dx}\Big |_{x=x_0 + \frac{1}{2}\Delta x} \approx \frac{T(x_0+\Delta x)-T(x_0)}{\Delta x},\nonumber\\
\frac{dT}{dx}\Big |_{x=x_0 - \frac{1}{2}\Delta x} \approx \frac{T(x_0)-T(x_0-\Delta x)}{\Delta x}.
\eea
If you combine the previous three equations, you'll have a nice approximation the second derivative of temperature in the x direction:

\be
\frac{d^2 T}{dx^2}\Big |_{x=x_0} \approx \frac{ T( x_0 + \Delta x ) - 2 T( x_0 ) + T( x_0 - \Delta x)}{(\Delta x)^2}
\ee
Now, if we want to move this approximation into python code, the trick is to remember that adjacent sites (i, i+1) are separated by \texttt{dx} or, $\Delta x$ in the language of the derivation.  This means that the second derivative at the ith site can be represented as:

\begin{verbatim}
ddT_dxdx = (temp[i+1][j][k] - 2.0*temp[i][j][k] 
   + temp[i-1][j][k])/(dx*dx)
\end{verbatim}

The other spatial derivatives can be similarly specified, see the example code for details.
%

\subsection{ What about the edges? }
If you think about it carefully, you'll realize that the above formula for  $\frac{d^2 T}{dx^2}$ won't work at the $x=0$ or $x=L_x$ edges of the chunk of iron.  Why?  Well, there is no $i+1$ node if you're at the uppermost, $i=N_x-1$, value, and similarly, there is no $i-1$ neighbor if you're at the $i=0$ location.  To get around this problem, we can use a different approximation.  Specifically, if we consider the simplest model of iron, in which the bar is completely isolated, there won't be a flow of heat to or from the outside, and so the flux would be zero across the edge.  This means that the derivative normal to an edge is zero.  The code that results (you can work it out from the above equations) is, 
\begin{verbatim}
# for the i direction:
if(i==0):     
  ddT_dxdx = (temp[i+1][j][k] 
    - temp[i][j][k])/(dx*dx)
elif(i==(Nx-1)):
  ddT_dxdx = (-temp[i][j][k] 
    + temp[i-1][j][k])/(dx*dx)
else:
  ddT_dxdx = (temp[i+1][j][k] 
    - 2.0*temp[i][j][k] 
    + temp[i-1][j][k])/(dx*dx)
\end{verbatim}

\subsection{ What about the time derivative?}
As a reminder, the heat flow equation is: 

\be
\frac{\partial T}{\partial t} = \kappa \left(\frac{\partial^2 T}{\partial x^2}+\frac{\partial^2 T}{\partial y^2}+\frac{\partial^2 T}{\partial z^2}\right)
\ee

We've worked out a way to approximate the right side of this equation, which describes the spatial variation in the temperature (or internal energy) of the solid.  The left side of the equation, $\frac{\partial T}{\partial t}$, describes the time rate of change of the temperature (or internal energy).  If we think about the movement of heat in the same way we've talked about the movement of particles in general mechanics, ie, if $v =\frac{dx}{dt}$, then $x(t+\Delta t) = x(t) + v\Delta t$., then the simulation can progress as follows.  Given an initial distribution of temperature, \texttt{temp[i][j][k]}, once can compute a change to that distribution, \texttt{d\_temp[i][j][k]}, and then say that the new energy distribution (after a time $dt$) is \texttt{temp[i][j][k]+d\_temp[i][j][k]}.  

\subsection{ Putting it all together}
With everything we've talked about, from an initial distribution, \texttt{temp[i][j][k]}, you should be able to figure out what the spatial changes are for every chunk of matter (\texttt{ddT\_dxdx}, etc).  Once these spatial changes are computed, you can figure out the change to each chunk's temperature (or internal energy), and then finally, if this change is added to each chunk, a new distribution of energy is obtained.  With this machinery, you should be able to simulate the temperature throughout the material for as long a time as you'd like.  

A fancier (ie, more accurate) simulation would include effects like: the transfer of heat from the bar to the surroundings by radiation, the temperature-dependent heat capacity of iron, variations in the composition of iron (via the inclusion of carbon in the steel) along the bar, or the ability to simulate quenching in oil or water, etc.

\subsection{
Heat flow from the forge, and from the rod 
}

The rate of energy flow per unit area, for something that's stuck into the forge is an unknown, and figuring out that value is one goal of this problem.  If we call this energy flow (in units of $\frac{W}{m^2}$), $P_{coals}$, then the effect on the system can be computed in the following way.  

If a certain mass element (chunk) is on the $k=0$ surface of the bar, the surface area touching the outside, $\Delta x \Delta y$ would permit the flow of energy.  Mathematically this means that the chunk of material has an additional \texttt{d\_temp[i][j][k]}, which we can compute:

\be
\frac{dU}{dt} = P_{coals}\Delta x \Delta y 
\ee

If we remember the definition of internal energy, $\Delta U = m c \Delta T$, we can solve for the additional temperature change, dT (or \texttt{d\_temp}), that occurs in the time interval $\Delta t$, because of the heat  flow from the coals.

\bea
\rho c \Delta x \Delta y \Delta z \frac{dT}{dt} &=& P_{coals}\Delta x \Delta y \nonumber\\
\frac{dT}{dt}\Delta t &=& P_{coals} \frac{\Delta x \Delta y \Delta t}{\rho c_p \Delta x \Delta y \Delta z } \nonumber\\
dT &=& +P_{coals}\frac{\Delta t}{ \rho c_p \Delta z}
\eea

Finally, the iron rod will likely be warmer than the rest of the blacksmithing room, and so there is a second flux, this time from the rod to the surroundings, that is given by the Stefan-Boltzmann Radiation law (for a blackbody).  According to this law, the power per unit area radiated by a hot object is given by, 

\be
P_{radiation}=\sigma T^4
\ee

where the power density is again in units of $\frac{W}{m^2}$, and where $\sigma=5.6705\times10^{-8}\frac{W}{m^2~K^4}$.  You should be able to use this relation to describe how heat leaves the surface of the rod which is not in contact with the coals. The specific implementation you could use (assuming again a mass element on the k=0 surface) is, 

\be
dT = -P_{radiation}\frac{\Delta t}{c_p \rho \Delta z}
\ee










\begin{thebibliography}{99}
\bibitem{pat_heller}
This is a common refrain heard informally in the Department of Curriculum and Instruction  at the University of Minnesota.
\bibitem{openmp}
It it likely that an implementation in a compiled language and some basic OPENMP parallel directives could speed this execution up significantly.
\end{thebibliography}
\end{document}